\documentclass[twocolumn,prd,preprintnumbers,amsmath,amssymb,noshowpacs]{revtex4}
\usepackage{graphicx}
\usepackage{dcolumn}
\usepackage{bm}
\usepackage{epsfig}

\begin{document}
\vspace{3 cm}

\title{Mass Limits on Neutralino Dark Matter}

\author{Rudy C. Gilmore}
\affiliation{Department of Physics, University of California, Santa Cruz, CA 95064}
\email{rgilmore@physics.ucsc.edu}
\date{\today}

\begin{abstract}
{We set an upper limit on the mass of a supersymmetric neutralino dark matter particle using the MicrOMEGAS and DarkSUSY software packages and the most recent constraints on relic density from combined WMAP and SDSS data.  We explore several different possible scenarios within the MSSM, including coannihilation with charginos and sfermions and annihilation through a massive Higgs resonance, using low energy mass inputs.  We find that no coannihilation scenario is consistent with dark matter in observed abundance with a mass greater than 2.5 TeV for a wino--type particle or 1.8 TeV for a Higgsino--type.  Contrived scenarios involving Higgs resonances with finely--tuned mass parameters can allow masses as high as 34 TeV. The resulting gamma--ray energy distribution is not in agreement with the recent multi--TeV gamma ray spectrum observed by H.E.S.S. originating from the center of the Milky Way.  Our results are relevent only for dark matter densities resulting from a thermal origin.}  
\end{abstract}

\maketitle

\section{Introduction}

The existence of cold dark matter in the universe is well--established by recent astrophysical observations.  While the particle nature of dark matter remains a mystery, its effect of promoting structure formation via hierarchical formation of gravitational potential wells from primordial density fluctuations is well--documented in both high resolution N--body simulations and observations of large--scale structure \cite{bertone-2005-405,2006astro.ph..9541P}.  The properties required for a thermally--produced particulate cold dark matter include high mass, lack of strong or electromagnetic couplings, and stability over the lifetime of the universe.  No standard model particle is capable of simultaneously satisfying these demands.  Minimal supersymmetry provides a number of viable candidates among its spectrum of fundamental particles \cite{primack-seckel-88,jungman-1996-267}.  The lightest of the four neutralinos, which in models we will consider is also the lightest supersymmetric partner or LSP, consists of various linear combinations of the neutral bino, wino, and Higgsino states.  The LSP has been proposed as an excellent candidate not only because of the above properties, but also because the expected cross section gives rise naturally to the observed dark matter mass density during thermal decoupling.  If supersymmetry is to solve the gauge hierarchy problem, it should be broken in such a way that the partner particles attain mass corrections on the order of the electroweak scale, $\sim 100$ GeV.  A weakly-interacting dark matter particle has the correct cross section to produce the observed relic density.  

Attempts to discover the particle nature of dark matter take on a three--pronged approach.  Terrestrial experiments to directly detect dark matter particles passing through the earth are underway \cite{sadoulet-2007-,leclercq-2006-,sanglard-2005-71}, and it is expected that the next generation of particle accelerators will be capable of producing a weakly interacting particle with a mass near the electroweak scale \cite{baltz-2006-74}.  In contrast to direct detection and production of dark matter, indirect detection techniques search for the products of dark matter annihilation.  Self-annihilations between clustered particles are expected to produce a variety of high-energy cosmic rays, photons, and neutrinos.  Sites for these events include the central density spikes of dark matter halos \cite{ullio-2001-64,bertone-2002-337,gondolo-1999-83,gnedin-2004-93,prada-barcomp,merritt-2004-92,merritt-2007-75,hall-2006-74}, diffuse radiation from the halo at large \cite{stoehr-2003-345,deboer-2007}, substructure within the galactic halo \cite{diemand-2006,pieri-2007}, and within astronomical bodies such as stars and planets, including our own earth and sun \cite{bergstrom-2000-,press-spergel-1985,halzen-2006-73}.  As mentioned in \cite{prada-barcomp}, the gamma ray flux per energy bin from neutralino annihilations in a region of space requires inputs from two factors, one being the cross section times the expectation value for the number of gammas produced, and the other an astrophysical variable determined by the distance and density profile of the region.  In addition to assuming a supersymmetric model, calculation of the signal from the center of our galaxy requires knowledge about the profile and normalization of the inner halo.  On scales larger than about 1 kpc, it is well--established through N--body simulations that the halo has a power--law radial density profile with index -1 to -1.5 \cite{navarro-1997-490,moore-1999,klypin-2000,power-2003-338}.  Frequent gravitational scattering events between dark matter and stars leads to an equilibrium profile of $\rho \sim r^{-3/2}$ in the inner 2 parsecs \cite{gnedin-2004-93}, where dark matter is a negligible proportion of the total mass.  Since the normalization is indeterminate, as a fiducial model we interpolate inwards from an NFW profile \cite{navarro-1997-490}.  While this may be a reasonable assumption, it is likely that baryonic compression, in which baryonic matter losing energy through radiative processes falls inward and consequently redistributes dark matter, has increased this number, possibly by several orders of magnitude \cite{prada-barcomp}.  For a weakly--interacting particle of mass greater than 100 GeV, it is plausible that a dark matter signal will be observed from the Galactic center by the current generation of ground--based gamma ray detection experiments, provided that baryonic compression has increased the central density by at least a modest amount over the fiducial value \cite{gnedin-2004-93}, although astrophysical processes could be a potentially troubling background for these searches \cite{zaharijas-2006-73}.       


A recent observation of a TeV gamma--ray signal from the Galactic center by the H.E.S.S. atmospheric Cherenkov telescope \cite{aharonian-2004-425,aharonian-2006-97} has motivated this determination of a theoretical upper limit on the mass of neutralino dark matter.  This result was confirmed recently by the MAGIC telescope \cite{albert-2006-638}.  The signal was also observed previously by the CANGAROO--II experiment \cite{tsuchiya-2004-606}, though this result was inconsistent with the newer H.E.S.S. results; the large flux observed by CANGAROO at low energies was not seen by H.E.S.S.  All groups have stated that the signal is consistent with a point source, and H.E.S.S. and MAGIC agree on a logarithmic slope of about -2.25.  Without a clear indication of an annihilating particle, such as line features in the spectrum, or an observation that disfavors an annihilation scenario, such as time variability, the source of the signal remains uncertain, although the extended power law nature of the observed spectrum does not fit well with the expected rollover shape of an annihilation spectrum.  An analysis by the H.E.S.S. group of their 2004 data, which extended to approximately 30 TeV, was unable to find any annihilation spectra which reproduced the observed power law, and they proposed that the signal must be primarily non--dark matter in origin.  While the newest data from H.E.S.S. does not seem to be consistent with the annihilation spectrum of either a supersymmetric neutralino or Kaluza--Klein dark matter particle \cite{bergstrom-2005-94}, models in which a signal from annihilating dark matter is masked by emissions of astrophysical sources are still possible \cite{aharonian-2006-97}.  Another analysis of the 2004 data by Profumo searched for optimal spectral fits based on final state channels; it was determined that the dark matter annihilation remains a possible interpretation of the H.E.S.S. data for a restricted set of final states \cite{profumo-2005-}.  The 2003 H.E.S.S. data, which extends to 9 TeV, can similarly be fitted with fewer constraints on the final states of the annihilating particle.  Mambrini and collaborators \cite{mambrini-2006-0601} searched for a neutralino annihilation spectrum in a non-universal supergravity model which could fit the 2003 data.  They were successful in finding reasonable fits, though none of the points in parameter space for their high mass candidates were consistent with WMAP constraints.  Possible astrophysical sources of TeV gamma--rays include jets or the shocks in the accretion flow into the central black hole \cite{yuan-2002-383,narayan-1998-492}.  Another possibility, which may be ruled out by lack of time variability, is a signal from particles accelerating near the event horizon of a rotating super--massive central black hole \cite{levinson-2000,aharonian-2004-425}.  

While various authors have conducted surveys of supersymmetric parameter space while categorizing dark matter candidates (e.g. \cite{baltz-2004-0410,baer-2005-0507,pierce-2004-70}), few papers have explicitly searched for upper mass limits on these candidates.  The effects of coannihilation in a low-energy effective MSSM model were considered in \cite{bednyakov-2002-66}, which calculated the degree to which relic density was reduced by various channels.  We conduct a similar survey, but with the goal of examining the regions of maximal mass in Higgsino coannihilation.  In \cite{edsjo-2003-0304}, the DarkSUSY software package was used to explore coannihilation in the mSUGRA parameter space.  In the chargino coannihilation region, the LSP is mostly Higgsino, and coannihilates with a nearly--degenerate chargino and second Higgsino.  This publication reported a cosmological limit of $\sim$1500 GeV resulting from chargino coannihilation, which is nearly consistent with our findings for a pure Higgsino in this region.  They also examined coannihilations of a bino--type LSP with sleptons and put forth the claim that coannihilation processes do not allow arbitrarily high masses, in contrast to some previous authors.  Higgsino dark matter in an mSUGRA framework was also considered in \cite{chattopadhyay-2006-632}, where a WMAP--favored mass range of approximately 1 TeV is found.  Wino--type dark matter appears in minimal anomaly--mediated supersymmetry breaking (mAMSB) models, and has a cosmological mass bound of over 2 TeV \cite{chattopadhyay-wino}.  Profumo \cite{profumo-2005-} looked at several different scenarios, including a mAMSB wino, annihilation through a heavy Higgs resonance, and QCD effects in gluino annihilations which can allow dark matter masses in the hundreds of TeV.

We use a model--independent approach in the calculation of the $m_{LSP}$ upper bound, one which could be applicable to many individual supersymmetry--breaking models.  To this end, we have used model inputs at the electroweak scale, allowing us to control individual inputs to the mass parameters and compute the dark matter relic density $\Omega_{LSP}$.  Calculations were done using the MicrOMEGAs \cite{belanger-2004-} and DarkSUSY \cite{gondolo-2004-0407} software packages.  Both of these codes compute SUSY relic density via a numerical solution of the Boltzmann equation, including the cross sections of any relevant coannihilation processes.  Both programs also provide several options for inputting the supersymmetric particle spectrum, including electroweak and GUT-scale minimal inputs, or determining the mass of each particle independently.  MicrOMEGAs calculates the contribution of individual decay channels to standard model products, as a function of the contribution of each channel to $\Omega^{-1}$, there being an approximate inverse relationship between the total cross section and final relic density.  DarkSUSY also provides the necessary tools to calculate the flux of high energy gamma--rays from halo annihilation, as well as a variety of other direct and indirect detection signals. 

\subsection{The Boltzmann Equation}

The density evolution of any particle $\chi$ in the thermal bath of the early universe is governed by the Boltzmann equation:
\begin{equation} \label{boltzdiff}
a^{-3}\frac{d(n_\chi a^3)}{dt}= \langle\sigma v \rangle \left( (n^{(0)}_\chi)^2 - n_{\chi}^2 \right ).
\end{equation}
Here $a$ is the cosmological scale factor, and $n$ and $n^{(0)}$ are the number density and equilibrium number density of the particle species.  The thermally-averaged cross section $\langle\sigma v \rangle$ must include all channels by which $\chi$ can interact, including coannihilation with other particles, in which the number densities of both species are important.  At some point in time the SUSY particle will no longer be able to remain in thermal equilibrium with its surroundings (``freeze-out'') and its co--moving number density will be nearly constant.  We limit ourselves to models which obey a discrete symmetry, R--parity, which prevents decays (but not two--body scattering) of SUSY particles into standard model particles \cite{jungman-1996-267}.  Any SUSY particle other than the LSP in existence at freeze--out will decay to the LSP state.  Following the derivation in \cite{dodelson}, the expression for the present relic density of a dark matter particle in terms of the cross section and freeze--out temperature is
\begin{equation} \label{bolt}
\Omega_{dm} h^2 \approx 0.3 x_f \sqrt{g_*} \frac {10^{-41} \mbox{cm}^2}{\langle\sigma v \rangle}.
\end{equation}
Here $x_f \equiv m_\chi/T_f$ where $T_f$ is the freeze--out temperature, and $g_*$ is the number of effective relativistic degrees of freedom of all species contributing to annihilation.  The important features here are that total annihilation cross section controls density, and that mass does not enter the equation except through a weak dependence in $g_*$ and $x_f$.

\subsection{Relic--Density Constraints}
The WMAP survey, when combined with recent observations of large--scale structure, currently provides the best constraints on the quantity $\Omega_{dm} h^2$, where $\Omega_{dm}$ is the ratio of dark matter density to the critical density $\rho_c = 1.88h^2 \times 10^{-29} \mbox{g cm}^{-3}$.  For our analysis, we use the most recent third year WMAP data combined with Sloan Digital Sky Survey large--scale structure data \cite{spergel-2006} to arrive at the tightest constraint on relic density, $\Omega_{dm} h^2=0.111^{+0.0056}_{-0.0075}$, here h $=0.709_{-0.032}^{+0.024}$ being the Hubble parameter in units of 100 km/s/Mpc.  We apply bounds equal to twice these 1$\sigma$ limits for the following analysis,
\begin{equation}
0.096 \leq  \Omega_{dm} h^2 \leq 0.122.
\end{equation}
These limits are sufficient to incorporate other recent measurements of $\Omega_{dm} h^2$ \cite{2006PhRvD..74l3507T,2006JCAP...10..014S}, which do not differ from our figure by more than $\sim 1\sigma$.  It should be noted that our results are not particularly sensitive to the relic density constraints, and there are larger sources of error involved in the calculation than the relatively minor variations in experimental determinations of $\Omega_{dm} h^2$.

\subsection{The Supersymmetric Neutralino}

The dark matter candidate we address in this paper is the lightest supersymmetric neutralino, denoted $\tilde{\chi}$, in the context of the minimal supersymmetric standard model (MSSM).  To express the neutralino mass states as a linear combination of Higgsino, bino, and neutral wino particle states, we diagonalize the mass matrix.
\begin{widetext}
\[ M_{\tilde{\chi}}^0 = \left( \begin{array}{cccc}
M_1 & 0 & -m_z \cos \beta \sin \theta_w & m_z \sin \beta \sin \theta_w \\
0 & M_2 & m_z \cos \beta \cos \theta_w & -m_z \sin \beta \cos \theta_w \\
-m_z \cos \beta \sin \theta_w & m_z \cos \beta \cos \theta_w & 0 & -\mu \\
m_z \sin \beta \sin \theta_w & -m_z \sin \beta \cos \theta_w & -\mu & 0 \\ 
\end{array} \right) \]
\end{widetext}

Here $\beta$ is the ratio of vacuum expectation values between the two Higgs doublets, $m_z$ is the mass of the $Z^0$, $\theta_w$ is the weak mixing angle, and $M_1$, $M_2$, $\mu$ are the U(1) and SU(2) gaugino and Higgsino mass parameters, respectively.  The physical states then become
\begin{equation} \label{neutmix}
{\tilde{\chi}_{i}}^0 = A_i\tilde{B} + B_i\tilde{W}^3 + C_i\tilde{H}_1^0 + D_i\tilde{H}_2^0, 
\end{equation}
with $A_i^2+B_i^2+C_i^2+D_i^2=1$.  Here $i=$ 1 to 4 is a particle index that will be suppressed in cases where the LSP is being discussed.  For our first models, we will be discussing instances in which the LSP is entirely Higgsino-- or wino--like, corresponding to the conditions $C^2+D^2 \approx 1$ or $B^2 \approx 1$, respectively.  The masses we are investigating are going to be strictly $>1$ TeV, and the off--diagonal blocks in the neutralino mass matrix are $< m_z$ so mixings are not large, and therefore the mass of the LSP is tightly controlled by the least of the three mass parameters.  The Higgsino parameter $\mu$ and SU(2) parameter $M_2$ also appears in the chargino mass matrix:
\[ M_{\tilde{\chi}^\pm} = \left( \begin{array}{cc} 
M_2 & \sqrt{2} m_w \sin \beta \\
\sqrt{2} m_w \cos \beta & \mu \\
\end{array} \right). \]
Again, the off diagonal parameters here are small compared to the mass scale of interest.  Thus our Higgsino-- and wino--type dark matter models come with a nearly--degenerate chargino (fermionic partner of charged Higgs and W bosons) built into the model at high energies.  This chargino will account for a large degree of the total annihilation cross section for these two dark matter types.  Incidentally, because we are interested in high--mass dark matter, $>1$ TeV, we will not be addressing the possibility of bino--type dark matter in this paper.  Because there is no degenerate chargino state in this case, the total annihilation cross sections and mass limits on bino dark matter tend to be much lower than the other two varieties, even with strong coannihilation from sfermionic particles.

\section{Mass Limit Results} 
\subsection{Pure States}
To explore cases in which the LSP is a pure wino or Higgsino, all SUSY masses are set to high values ($>10$ TeV) except for either the Higgsino or SU(2) (wino) mass parameter.  The other relevant particle which appears in this situation is a slightly heavier chargino, and in the Higgsino case a second nearly degenerate Higgsino--type neutralino.  As the other particles are at a significantly higher mass scale, they will be thermally suppressed prior to dark matter freeze--out and will not affect relic density.  Our results for this particular region show a monotonically increasing relic density with increasing mass, with no dependence on the $\tan\beta$ parameter.  Our bounds from combined WMAP and SDSS for these cases are:
\begin{eqnarray}
0.99 & \leq & m_{\tilde{\chi}} (\mbox{TeV}) \leq  1.12 \textrm{ (Higgsino) }  \\
2.10 & \leq & m_{\tilde{\chi}} (\mbox{TeV}) \leq  2.38 \textrm{ (wino) }
\end{eqnarray}

Our mass limit for a pure wino state is consistent with that mentioned by Profumo \cite{profumo-2005-}, who quoted a function $\Omega_{dm}h^2 = c(m_{\tilde{\chi}}(\mbox{TeV}))^\gamma$ with $0.0225 \leq c \leq 0.0255$ and $1.90 \leq \gamma \leq 1.92$.

\subsection{Coannihilation with a Sfermion}
In ordered to systematically test the effects of coannihilation with a Higgsino, we tested each sfermion mass parameter, originally set to high values, by shifting them down to the coannihilation region, to a mass $m_{ca}$ which is slightly larger than $\mu$.  Beginning with the limits set by Higgsino--chargino coannihilation, we attempt to find regions where these processes allow a larger Higgsino mass by increasing the effective cross section for annihilation.  Depending on the specific interaction strengths for processes involving this new particle, the relic density may be increased or decreased as $m_{ca}$ is brought lower, that is, coannihilation may have a positive or negative effect on the mass limit.  In our low--energy effective supersymmetric model the sfermion masses are all free parameters.  For our notation we write $m_{\tilde{q}i}$, $m_{\tilde{u}i}$, $m_{\tilde{d}i}$, $m_{\tilde{l}i}$, and $m_{\tilde{e}i}$, with i = 1,2,3 being the generation index, for the left--handed squark doublet, right--handed up and down singlets, left--handed slepton doublet, and right--handed slepton singlet, respectively.  This gives 15 free mass parameters to examine in this coannihilation calculation.

\begin{figure*}[htpb]
\centering
\epsfig{file=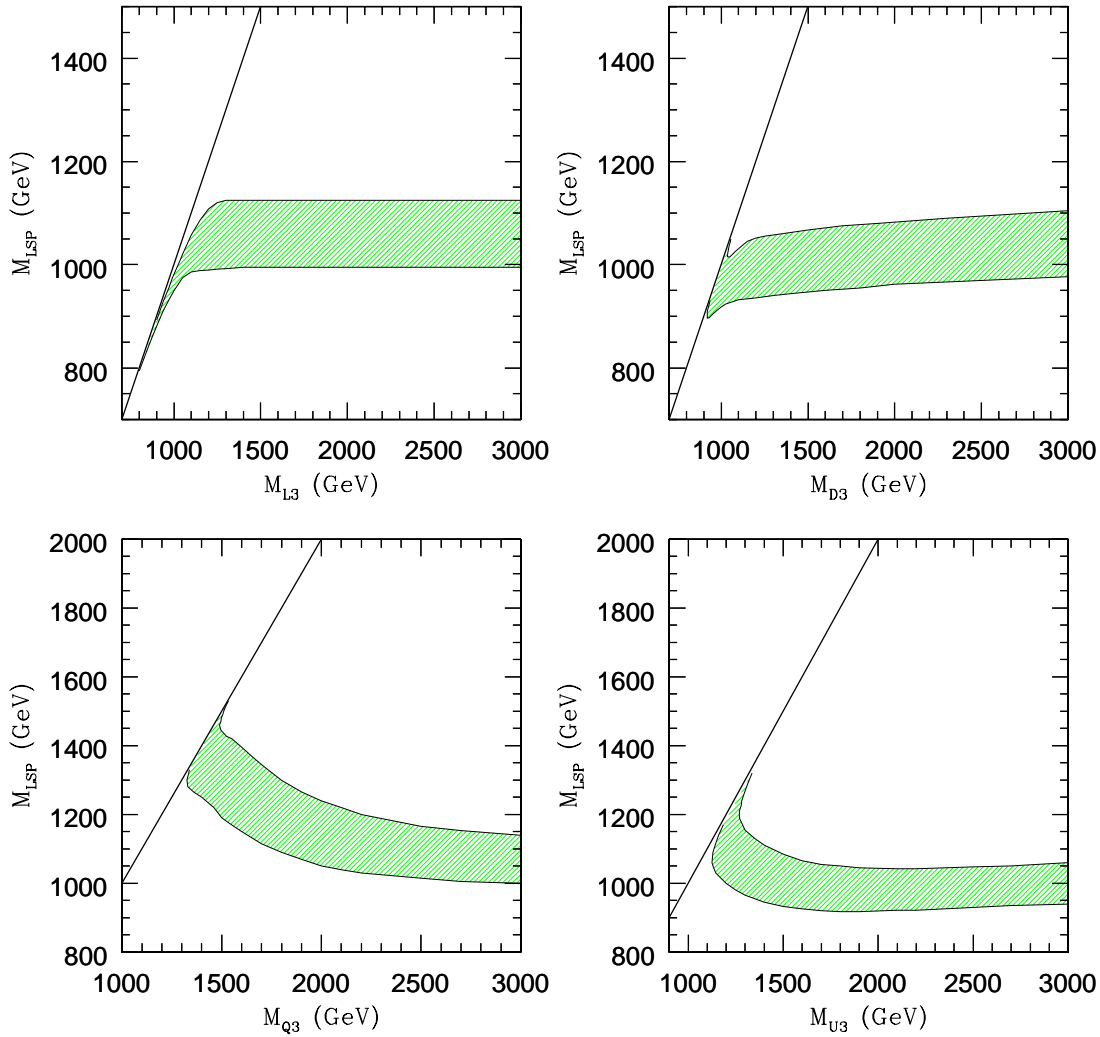, width=15cm}
\caption{Results for Higgsino--squark coannihilation.  We show allowed regions in $m_{\tilde{f}}$-$\mu$ space, where $m_{\tilde{f}}$ is the mass for 4 different sfermion species shown here: $\tilde{q}3$, the left-handed doublet of 3rd--generation squarks in the lower left, $\tilde{u}3$ and $\tilde{d}3$, the right--handed singlet partner of the top and bottom quarks in the lower and upper right, and $\tilde{l}3$, the left-handed doublet of 3rd--generation leptons in the upper left.  The two lines indicate the upper and lower $2\sigma$ bounds on relic density.  The physical masses of the sfermions in each case are very nearly equal to our mass parameter, with corrections $<5$ GeV that can be ignored.  Thus, the contour lines must end at $m_{\tilde{f}} = \mu$, for the Higgsino to remain the LSP.  The highest vertical point in the regions for $\tilde{u}3$ and $\tilde{q}3$ indicate the highest possible Higgsino mass.  In the case of $\tilde{l}3$, the effect of coannihilation on the upper mass bound is strictly negative.  The $\tilde{d}3$ parameter has more complex behavior, but the overall effect is negative.}
\label{fig:coannihilation_contours}
\end{figure*}

Using MicrOMEGAs and confirming our results with DarkSUSY, we determine that only coannihilation with the third generation of squarks allows the Higgsino mass to be increased beyond the amounts in the previous section while conforming to experimental limits.  This is shown graphically here in Figure \ref{fig:coannihilation_contours}, which show regions in $m_{ca}$--$\mu$ space that fall within the $2*1\sigma$ bounds on relic density.  For these figures, all masses other than $\mu$ and $m_{ca}$ are set to 50 TeV.  At this mass these particles are effectively removed from the early--universe Boltzmann equation, as their number densities are exponentially suppressed prior to freeze--out.  The ratio of Higgs vacuum expectation values $\tan\beta$ is set to 30, and the supersymmetry breaking parameters $A_i$ are set to 0.  No other sfermion increases the maximum Higgsino mass when allowed to coannihilate.  Also, the effect of coannihilation with two or more sfermions is not found to be cumulative in general.  When $m_{\tilde{q}3} \approx \mu$, bringing down any other mass either increases $\Omega_{dm} h^2$ or has a negligable effect.  We did not find any cases in which compound coannihilation with several sfermions had a substantially effect on the mass bound.

The ratio of Higgs doublet vacuum expectation values $\tan \beta$ has a significant effect in the sfermion coannihilation region.  Our previous analysis was done with a typical value of $\tan \beta = 30$, but, with a higher value, the effects of coannihilation are increased.  Third generation sfermion right--hand singlet parameters u3 and d3 can also have a slight enhancement effect in this regime.  The optimal combination is found to be $\mu \approx m_{\tilde{q}3} = 0.9 m_{\tilde{u}3} = 0.9 m_{\tilde{d}3}$.  In this case, $m_{\tilde{\chi}}$ may grow as large as 1.8 TeV while remaining within experimental density bounds.  This is the greatest mass which was possible under any combination of coannihilating sfermions in our Higgsino scenario.

For a wino--type LSP, no sfermion coannihilation arrangement was found to have the effect of raising the mass bound; all sfermion coannihilation schemes cause the mass limit to decrease.  Therefore the highest mass limit for this type of dark matter is that found in the previous section.  
 
\subsection{Annihilation through a Massive Higgs Resonance}

Another mechanism through which the annihilation cross section of a neutralino LSP might be greatly enhanced is via a heavy Higgs resonance, $m_A\approx 2m_{LSP}$.  At the multi--TeV scale, the masses of the heavy CP--even Higgs, CP--odd Higgs, and charged Higgs are all approximately degenerate. 

\[
m_A \approx m_H \approx m_{H^\pm}
\]

Under this arrangement, not only is the cross section of LSP annihilation enhanced by the resonance, but so are coannihilations between any nearly degenerate charginos or next to lightest neutralinos.  It is expected that the cross section will be dominated by the CP--odd Higgs channel, as the contribution from CP--even Higgses vanish in the low velocity limit due to the requirement of CP conservation in the intermediate state \cite{jungman-1996-267}.  Profumo \cite{profumo-2005-} analyzed this region of parameter space in minimal supergravity (mSUGRA) and anomaly--mediated (mAMSB) SUSY breaking models with non-universal Higgs masses.  He found upper LSP mass limit of approximately 5 and 12 TeV for mSUGRA and mAMSB, respectively, utilizing $2\sigma$ WMAP bounds and $\tan \beta = 40$.  We have followed a similar program, taking the low--energy neutralino and chargino mass matrix inputs as free parameters, with the aforementioned constraint on the LSP mass, therefore exploring over the vector space of neutralino and chargino mixings.  For two multi--TeV neutralinos interacting at zero velocity, the cross section for annihilation through a CP--odd Higgs is \cite{griest-1991,profumo-2005-},
\begin{equation}
\langle \sigma v \rangle = \frac{g_{A \tilde{\chi}\tilde{\chi}}^2}{8\pi\Gamma_{A}^2} \sum_f c_f |g_{Aff}|^2 \approx \frac{2 \pi g_{A \tilde{\chi}\tilde{\chi}}^2}{m_{\tilde{\chi}}^2 \sum_f c_f |g_{Aff}|^2}
\end{equation}
where $g_{A\tilde{\chi}\tilde{\chi}}$ and $g_{Aff}$ are the vertex factors for the coupling between the Higgs and neutralino and final--state fermion species $f$, respectively, and $\Gamma_{A}$ is the Higgs width.  The vertex factors appearing in a neutralino--Higgs or chargino--Higgs junction involve products of gaugino and Higgsino mixing factors, and are therefore sensitive to the exact choice of mass parameters.  For two neutralino LSPs annihilating through a CP--odd Higgs, we have \cite{edsjo-1997-} 

\begin{eqnarray}
& g_{A \tilde{\chi}\tilde{\chi}} = (gB-g'A)(C\sin\beta-D\cos\beta) \\ \vspace{2 mm}
& g_{Auu}=\frac{gm_{u}\cot\beta}{2m_W} \\  \vspace{2mm}
& g_{Add}=\frac{gm_{d}\tan\beta}{2m_W}
\end{eqnarray}

where the LSP composition is denoted by parameters A through D as in equation \ref{neutmix}.  Here `u' refers to up--type quarks and neutrinos, and `d' refers to down type quarks and charged leptons, and g and g' are the SU(2) and U(1) coupling constants.  Because the $g_{A \tilde{\chi}\tilde{\chi}}$ vertex factor is determined by products of gaugino and Higgsino fractions, the largest factors and highest annihilation enhancements tend to occur when the neutralino is an even mixture of gaugino and Higgsino particle eigenstates.  As derived in \cite{profumo-2005-}, the maximum cross section for the CP--odd Higgs channel is at $\tan\beta \approx \sqrt{m_t/m_b} \approx 6.4$; our investigation of different values for $\tan\beta$ found the highest mass limits at approximately this point.

\begin{figure*}[htpb]
\centering
\epsfig{file=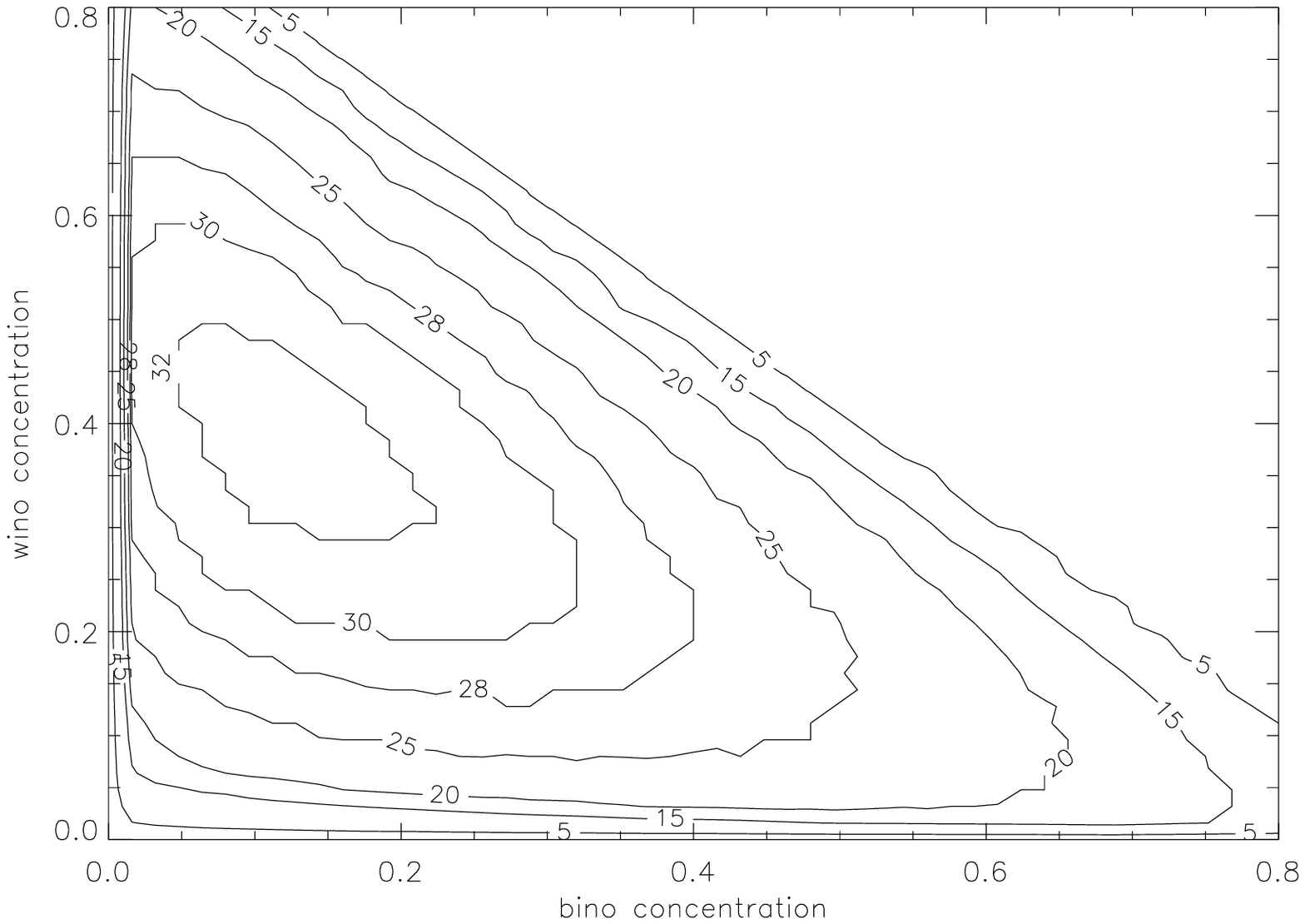, width=10 cm}
\epsfig{file=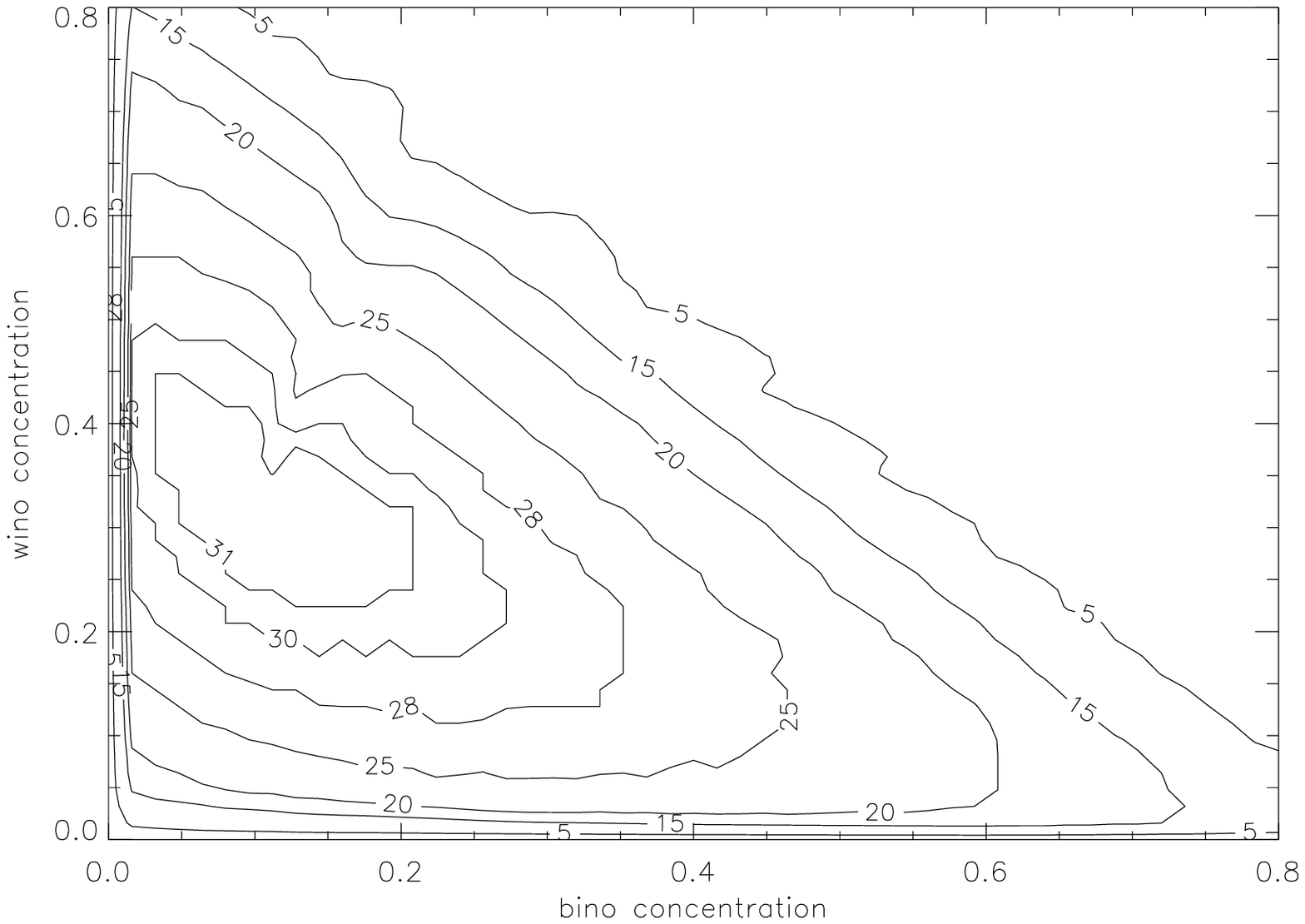, width=10 cm}
\epsfig{file=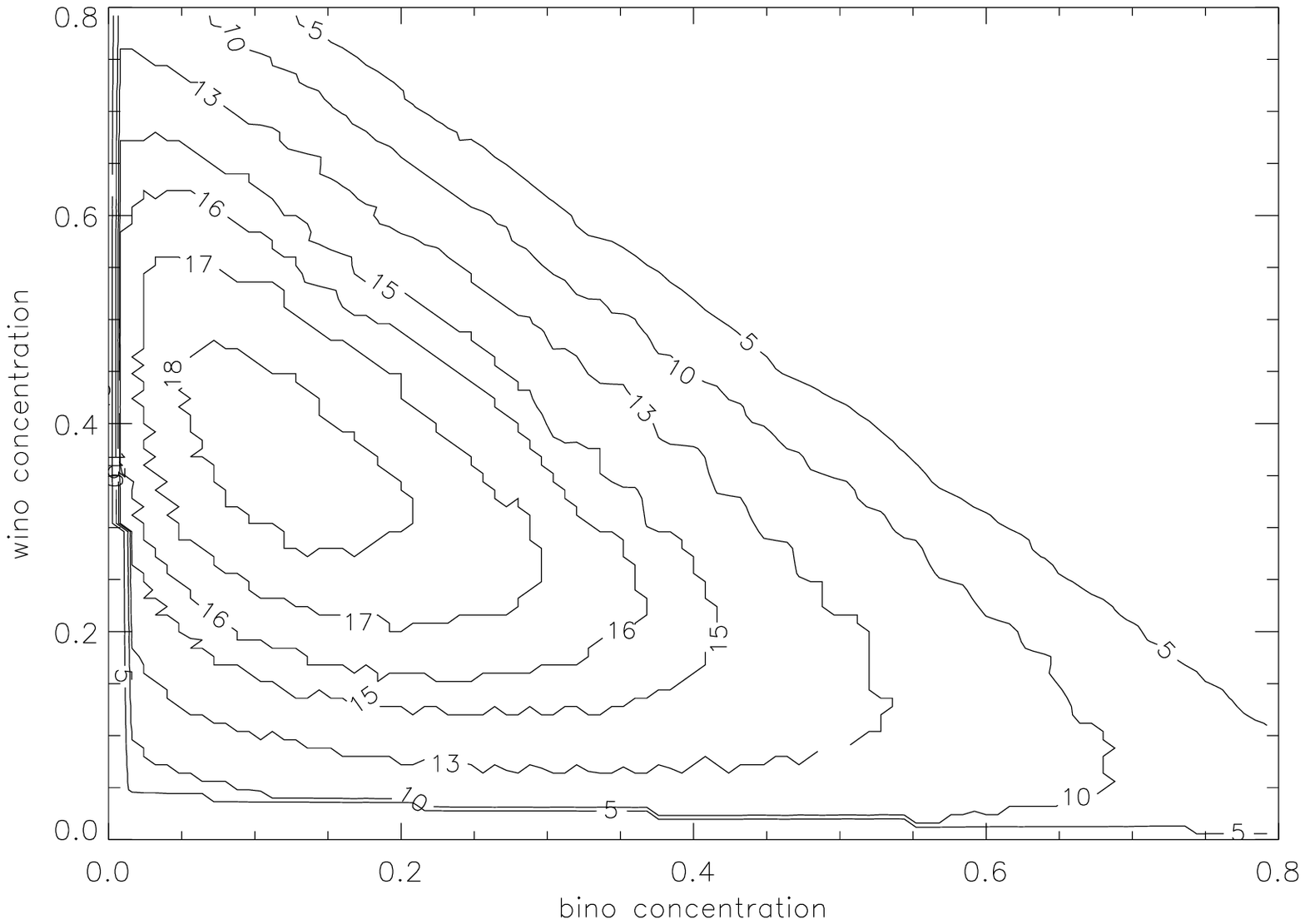, width=10 cm}
\caption{Resonant heavy Higgs annihilation mass limit plots for $\tan \beta$= 2 (top), 10 (middle), and 50 (bottom).  The axes are the fractional wino and bino states of the LSP, with the Higgsino fraction being the remainder.  The contours show the upper mass bounds in TeV.}
\label{fig:schan}
\end{figure*}

Our results for this type of model can be seen in Figure \ref{fig:schan}.  We systematically calculated mass limits corresponding to the upper $2*1\sigma$ experimental bound over a large number of neutralino mixtures.  This was done by setting $m_A = 2\mu \approx 2m_{\tilde{\chi}}$, and scanning over values of $\mu-M_1$ and $\mu-M_2$.  These differences alter the particle state content of the LSP via the neutralino mass matrix.  The final results are shown in terms of the fractional LSP composition.  As $\tan\beta$ is also an independent, relevant parameter, we included this as a variable and searched from $\tan\beta$ = 2 to 50.  The results from this scan are plotted as mass contours as a function of fractional neutralino composition.  We show plots for 3 values of $\tan\beta$ in Figure \ref{fig:schan}.  

\section{Detection Rates}
For each of the model types thus discussed, we have calculated the resulting local gamma--ray flux from annihilations in the galactic center.  For our halo model, we utilized the predictions of \cite{gnedin-2004-93}, together with their fiducial value for the density normalization at the maximum radius of the central black hole's sphere of influence, $\sim$ 2pc.  This factor is highly uncertain and it is possible that it has been increased by baryonic infall.  Dark matter annihilation can also have an effect on the density profile of the innermost part of the cusp, changing the signal intensity by a multiplier refered to as the boost parameter \cite{bertone-2005-72}, an effect not considered here.  A reader wishing to use a different normalization can simply raise or lower our flux measurements by the square of the normalization multiplier, or by the boost factor in the case of modifications to the inner halo profile. 

The total cross sections for all annihilations to continuous and line features in the spectrum were calculated using the DarkSUSY program.  In order to show a typical spectrum as seen by a detector, we have folded the distribution with a gaussian with an energy width of 15 percent, as was done in \cite{bergstrom-2005-95}.  This also allows a reasonable visual comparison to be made of the prominence of the line emission feature against the continuous output.

\begin{figure*}[htpb]
\centering
\epsfig{file=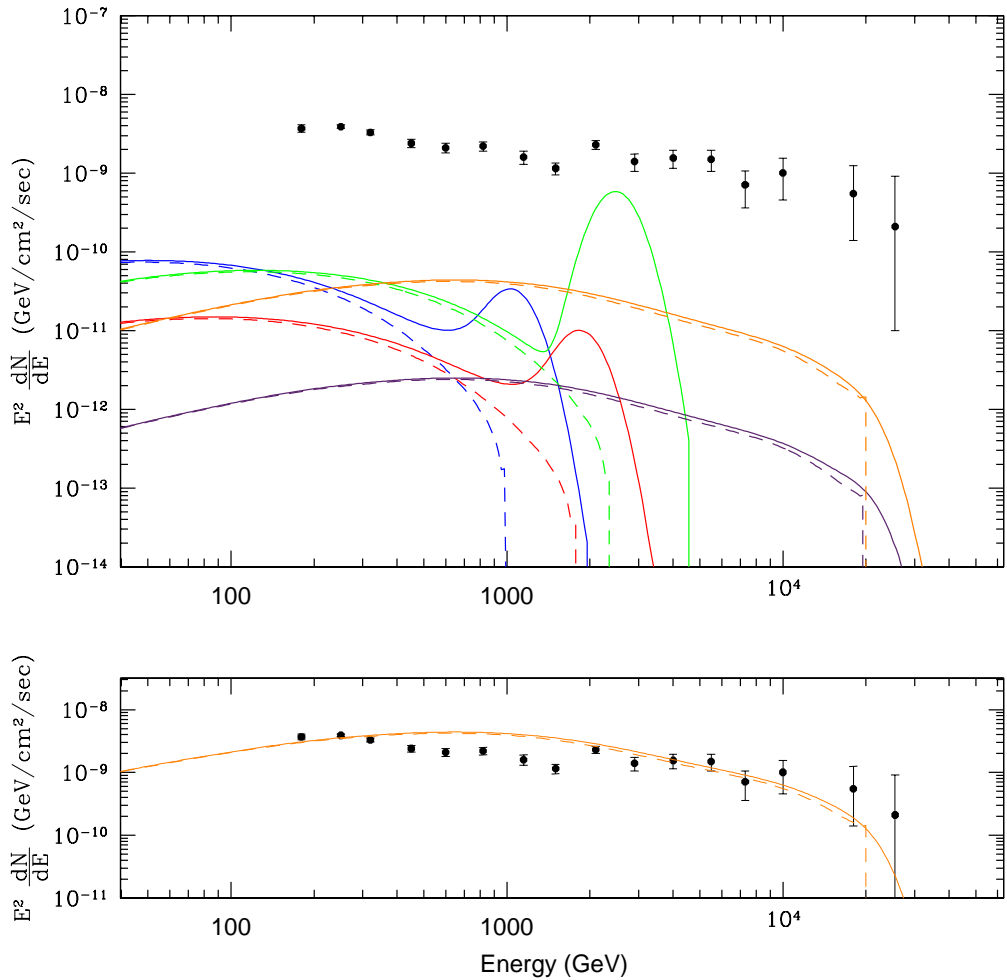, width=14 cm}
\caption{In the upper plot, we summarize our findings by showing the resulting local gamma--ray flux from the galactic center in several annihilation scenarios using the halo model of \cite{gnedin-2004-93} with fiducial normalization (no baryonic compression), and compare to the latest observations of the H.E.S.S. experiment (black data points, \cite{aharonian-2006-97}). The dashed lines show the true continuous distribution, while the solid lines show the total (continuous plus discrete) emission spectra as seen by a detector with an energy resolution of 15 percent.  The blue line is a 1 TeV Higgsino, coannihilating with a nearly degenerate chargino and second Higgsino.  The red line shows the same model with coannihilation from a 3rd generation squark, at a mass of 1.8 TeV.  The green line is a 2.4 TeV wino.  The purple and orange lines are both a mixed type neutralino annihilating through a heavy Higgs resonance.  The orange model has been optimized by fine tuning of the resonance, so that the cross section and resulting flux are maximized, while the purple line shows a more typical model.  The lower plot demonstrates an attempt to fit a Higgs resonance model to the H.E.S.S. data.  A factor 10 density boost is applied, resulting in a $10^2$ increase in flux above the fiducial value.}
\label{fig:dmodflux}
\end{figure*}

In Figure \ref{fig:dmodflux} we show the gamma ray flux from 5 different models, corresponding to the scenarios we have discussed.  Clearly, with the fiducial halo normalization none of these annihilation models can account for a significant fraction of the flux observed by the H.E.S.S. telescopes and confirmed by MAGIC.  However, the highly uncertain contribution to the central density from baryonic compression has not been taken into account in these results.  As flux increases quadratically with particle density, a rather modest compression factor of 10 would increase flux by 2 orders of magnitude, enough to bring our more strongly annihilating models to the levels observed by these atmospheric Cherenkov telescopes.  However, the H.E.S.S. data also maintains an approximate power--law profile for 2 logarithmic decades, something that none of our models can reproduce even with a carefully adjusted density normalization.  Even the most massive particles we found to be capable of satisfying relic density constraints exhibit a roll--off behavior that is not in the observed spectrum.

Two plots for the annihilation through a heavy Higgs resonance are shown.  We have chosen a mass of 20 TeV here, the heaviest neutralino for which a cross section could be computed without resorting to extrapolation in certain DarkSUSY routines.  Because mean halo velocities are much lower than those at freeze--out, annihilation cross sections are more sensitive in the halo to the exact relation between $m_A$ and $m_{\tilde{\chi}}$ as there is little smearing out of the center--of--mass energy due to thermal velocities.  Thus models that yield similar relic densities can have very different halo annihilation cross sections.  To illustrate this we have displayed both an optimized model (orange) with $m_A = 2m_{\tilde{\chi}}$ and a more typical model (purple) where the relation is only approximate. 

\section{Discussion}

Higgsino-- and wino--type dark matter annihilate much more strongly than bino dark matter at a given mass, and are capable of satisfying relic density constraint with masses at the TeV scale, something that is not possible with a bino particle, even including sfermion coannihilations.  Primarily Higgsino-type dark matter can arise, for example, in the `focus point' region of minimal supergravity (mSUGRA) \cite{feng-2000-84}.  In this model, the supersymmetric scalar quarks and leptons are set to high masses by a single GUT scale parameter, and the neutralino becomes a Higgsino-bino mixture.  As this scalar mass term is increased, the neutralino LSP becomes more Higgsino--like, and features an increasing cross section \cite{edsjo-2003-0304,feng-2000-482}.  Since the highest mass LSPs in the focus point region are nearly pure Higgsino, we have constrained this and other Higgsino-type models by considering this limiting case.  Dark matter which is predominantly wino--type appears in minimal anomaly--mediated supersymmetry breaking \cite{feng:1999hg}.  MicrOMEGAs provides information on the annihilation channels relevant to the total LSP cross section in a particular model.  For the case of Higgsino--chargino coannihilation, a large number of channels provide modest contributions to the total cross section.  Chargino and neutralino annihilation to quarks are the most important processes, with annihilations to leptons and gauge bosons, and double chargino annihilation to quarks being the other relevant channels.  In the $>$TeV mass range of interest, even the heaviest standard model particles are essentially massless, and there is little difference in available phase space between different generations, hence little variation in cross section.  Not surprisingly, there is also relatively little variation in channel contributions as the mass scale and resulting relic density are altered.

The introduction of a coannihilating squark opens the possibility of tree--level annihilations to gluons.  Annihilations to gluons, as well as coannihilations between Higgsinos and squarks to quarks and gauge bosons, are the major new channels available.  The decreased mass of a given sfermion increases the t--channel amplitude for scattering from neutralinos to fermions, but only for that particular flavor of the sfermion. These same vertices explain why the third generation of squarks are unique as the only effective set of coannihilating partners.  The appearance of the corresponding quark mass terms limits these channels to cases involving heavy quark masses.  The quantity $\beta$ appears in several vertex factors involving Higgs and Higgsino iterations with matter \cite{edsjo-1997-}.  Processes with a factor of $\tan\beta$ or $(\sin\beta)^{-1}$ in the amplitude include chargino and neutralino annihilation with a sfermion to a gauge boson and fermion, and these are primarily responsible for the decrease in relic density in scenarios with squark coannihilation with increasing $\tan\beta$. 

An effect which is relevant for Higgsino and wino models is the Sommerfeld enhancement, appearing in the context of weakly--interacting non--relativistic particles with a mass much greater than the W--boson.  This non--perturbative effect has been shown to decrease relic abundance by as much as 50$\%$ for a wino--type LSP and 10$\%$ for a Higgsino--type \cite{hisano-2007-646}.  These authors study wino dark matter as an example and find upper WMAP bounds of 2.7 to 3 TeV.  Neither DarkSUSY nor MicrOMEGAs accounts for this effect; in addition to these considerably higher new bounds for wino dark matter, our bounds involving pure Higgsino dark matter could be enhanced by a slight amount.   

When annihilation through a heavy Higgs s--channel resonance is considered, allowed masses can go into the tens of TeV.  The examination of this scenario was done using the DarkSUSY software.  As expected, the neutralinos which had the largest cross section were approximately even mixtures of Higgsino and gaugino states.  From Figure \ref{fig:schan}, it is clear that there is little change in the general topography of the relation between neutralino mixture and mass bounds with changing $\tan\beta$.  The maximum mass limits do change as a function of $\tan\beta$, rising a small amount from 32 TeV at $\tan\beta = 2$ to about 34 TeV at $\tan\beta =$ 5 to 8, and then decreasing from that point down to 18 TeV at $\tan\beta = 50$.

Our results with MicrOMEGAS were tested against the DarkSUSY code, and the programs were found to generally be in agreement over the parameter space of interest, except in the case of annihilation through a Higgs resonance.  For our calculations involving pure Higgsino and wino states, the difference in relic density calculations were no larger than 2.5 percent, and in certain cases where sfermion coannihilation was considered it was no greater than 9 percent.  It was noted during while investigating sfermion coannihilation that the two codes produced highly disparate results in certain situations.  These problems were determined to be a error in the DarkSUSY software that only appears at mass scales higher than we have considered here, and did not appear to be an issue for our results in the coannihilation region.  For the Higgs resonance models, there was a significant difference in the predictions of the two codes, sometimes by as large as a factor of 2.  The DarkSUSY results, which we have presented, tended to output higher mass limits than MicrOMEGAs.  While this does mean that our results in this area should be taken only as approximate, our conclusion that this scenario is unlikely and cannot explain current gamma ray observations is not altered.   

While there is no concrete upper bound on the scale of supersymmetry breaking, a mass well into the TeV range is certainly disfavored by constraints from gauge coupling unification \cite{ellis-1992-287,bourilkov-2005-20}.  Another `absolute' bound comes from partial wave unitarity, which provides an upper limit on the mass that any thermally produced dark matter particle can have, by placing a constraint on the cross section in equation [\ref{boltzdiff}].  This bound is applicable as long as the annihilation cross section arises primarily from s--wave terms.  The mass limit set by unitarity is $\Omega_{dm}h^2 \geq 1.7*10^{-6}\sqrt{x_f}(m_{dm}/TeV)^2$ \cite{griest-kamionkowski}, which leads to a maximum relic density of about 120 TeV.  While this mass is well above that of any of the MSSM models we examined, it may become important when considering thermally--produced heavy dark matter candidates from other particle physics extensions to the standard model.  However, it should be noted that the unitarity bound can be violated in the case of a strong resonance, in which the assumption of s--wave dominance breaks down \cite{hui-2002}.  Another case in which the unitarity bound is not applicable comes from possible non--perturbative factors in the annihilation cross section which could affect heavy ($>$500 GeV) Higgsino-type neutralinos \cite{hisano-2005-71}.  These factors appear only at low velocity and would not affect the physics of the dark matter during freeze--out but could thus affect halo interactions, greatly increasing flux levels from annihilations.

The mass limits we have set in this paper apply only to neutralino dark matter in the MSSM model which attains a relic density through thermal freeze--out in a standard cosmology.  We can make no claims about cases in which the dark matter is produced through non--thermal processes.  These could include scenarios in which the dark matter is produced non--thermally, possibly by a late--decaying scalar field \cite{gelmini-2006-74}, or one in which entropy is produced after freeze--out.  This latter case could happen for a variety of reasons (\cite{jungman-1996-267}, for review) and would have the effect of violating the standard assumption of constant comoving entropy density, which would reduce relic density.  The unitarity bound would not apply in this situation, as a very massive dark matter particle with ordinary cross section could still attain the correct density today.  

\section{Summary}
We have determined the masses of pure Higgsino-- and wino--type thermally--produced dark matter which are consistent with the latest density constraints on dark matter, defined here as twice the 1$\sigma$ bound determined by combined SDSS and WMAP--3 data.  In the absence of any coannihilation processes with scalar fermions, the suitable mass range is found to be between 0.99 and 1.12 TeV for a pure Higgsino and 2.10 and 2.38 for a pure wino state.  Coannihilation with partners of the 3rd--generation quarks is found to increase this limit modestly for Higgsino type dark matter to an upper limit of about 1.80 TeV, with fine tuning in the mass parameters and $\tan\beta$, but no coannihilation model can increase the mass limit for a wino--type particle.  Allowing the dark matter to exist as a bino pure or mixed state tends to sharply decrease mass limits, and bino mass limits were always found to be in the sub--TeV range.  The other class of models which we examined utilized annihilation of the LSP through a heavy Higgs resonance.  Viable models with LSP masses as high as 34 TeV were found, though these scenarios are sensitive to the both the neutralino mixture and the resonance condition $m_A = 2m_{\tilde{\chi}}$, and are therefore dependent on fine--tuning.  A computation of the VHE gamma ray spectrum which could be observed with an atmospheric Cherenkov telescope showed that even the largest masses we found are not adequate for fitting the observed H.E.S.S spectrum.  This observed event rate is also considerably higher than our predictions, although this is not necessarily a problem because of uncertain normalization of the dark matter profile in its innermost regions.

The author wishes to thank Joel Primack, Patrick Fox, Andreas Birkedal, Gordon Kane, and Stefano Profumo for advice and guidance.  This project would not have been possible without the MicrOMEGAs and DarkSUSY teams, and in particular Fawzi Boudjema and Genevieve Belanger from the former, and Paolo Gondolo, Joakim Edsjo, and Edward Baltz from the latter, who provided assistance on various issues.  This work was supported in part by NSF grant AST--0607712 and by the University of California, Santa Cruz Division of Graduate Studies.

\end{document}